# Trapping of slow-speed particles in a gas cell by the nonhomogeneous electromagnetic field intensifying with time

Azad Ch. Izmailov

*Institute of Physics, Azerbaijan National Academy of Sciences, Javid av. 33, Baku, Az-1143, AZERBAIJAN*

*email*: azizm57@rambler.ru

## *Abstract*

Author suggests and analyzes new universal trapping method of comparatively slow-speed particles of a rarefied gas medium in the potential well induced by the nonhomogeneous electromagnetic field increasing with time (up to some moment). Given method is especially effective at inelastic collisions of particles with walls of the gas cell when necessary preliminary slowdown of particles is possible for their following capture even to a highly shallow potential depth. Corresponding sufficiently compact and simple electromagnetic traps may be used for capture and accumulation not only slow-speed micro- and nano-particles in the high vacuum but also atoms and molecules of a rarefied gas in a cell.

## 1. Introduction

Electromagnetic traps of micro- and nano-particles in the high vacuum open new possibilities for contactless measurements of forces acting on given particles with extremely high accuracy and also allow micromanipulations of such particles, for example, in hollow waveguides [1]. Even more important is the development of effective methods of trapping and localization of sufficiently slow-speed atoms and molecules, in particular, for ultrahigh resolution spectroscopy [2] and for creation of more precise standards of time and frequency [3].

In the present work we consider new sufficiently universal mechanism of trapping of free particles (in particular atoms and molecules) in a gas cell by the nonhomogeneous electromagnetic field increasing with time (up to some moment). Efficiency of such trapping essentially increases in cases of inelastic reflection of given particles from walls of the gas cell. Then necessary preliminary slowdown of particles is possible for their following capture to the electromagnetic trap. Probability of such "cooling" collisions increases with time and therefore accumulation of trapped particles occurs during growth even a highly shallow potential well of the trap. Depending on whether particles have electric (magnetic) moment, it is possible to use the controllable electric (magnetic) field or laser radiation for their trapping. For visual demonstration of the proposed method, at first we will analyze the comparatively simple one-dimensional model of a cell and the electromagnetic trap with a gas of structureless particles (section 2). Then we will discuss possible generalization of obtained results on real two- and three-dimensional systems and also on atoms and molecules of rarefied gases (final section 3).

## 2. One-dimensional model of the trap

Let us consider a collection of noninteracting identical point-like particles, which are under conditions of ultrahigh vacuum between plane-parallel walls of the one-dimensional cell with coordinates $x = \pm l$. We suppose that controllable electromagnetic field is superimposed on this cell, which creates the potential well for a particle described by the following function $U(x,t)$ of time $t$ and coordinate $x$ ($-l \leq x \leq l$):

$$U(x,t) = 0.5\, m\, \omega(t)^2 x^2, \qquad (1)$$

where $m$ is the particle mass and $\omega(t)$ is the positive value nondecreasing with time $t$. Corresponding motion equation of a particle in the well (1) is characteristic for the one-dimensional oscillator with the changing frequency $\omega(t)$:

$$\frac{d^2 x}{dt^2} + \omega(t)^2 x = 0. \qquad (2)$$

Let us consider any particle, which reflects with a velocity $dx/dt = v_1$ from the cell wall with the coordinate $x = -l$. For example, we present the following time dependence $\omega(t)$:

$$\omega(t) = \omega_0 + \sigma \cdot t \cdot \eta(T - t) + \sigma \cdot T \cdot \eta(t - T), \qquad (3)$$

where $\omega_0$ is the value $\omega(t)$ in the initial moment $t=0$, $\sigma > 0$ is the constant parameter, $\eta(y)$ is the step function ($\eta(y) = 1$ if $y \geq 0$ and $\eta(y) = 0$ when $y < 0$), and $T$ is the moment when growth of the frequency $\omega(t)$ stops and further $\omega(t) = (\omega_0 + \sigma \cdot T)$ is the constant. Fig.1 shows the example of the particle motion $x(t)$, calculated on the basis of the Eq.(2) with corresponding boundary conditions, at the change of the potential well (1) according to the time dependence (3). We see that, after reflection of the particle with a comparatively small velocity $v_1$ from the cell wall (with the coordinate $x = -l$), this particle does not reach the opposite wall (with the coordinate $x = l$) and turns back (in the point $z$ in Fig.1) because of intensifying (with time) gradient force, which is directed to the center $x = 0$ of the potential well (1). Following motion of the particle is oscillations with the increasing frequency and decreasing amplitude up to the moment $t=T$, after which the value $\omega(t)$ (3) is constant. Then the particle undergoes usual harmonic oscillations (with the constant frequency and amplitude) already without collisions with cell walls (Fig.1). Thus it is possible to realize the trapping of sufficiently slow-speed particles by the nonhomogeneous electromagnetic field increasing with time (up to the definite moment). It is obvious, that such a trapping will not be possible for comparatively fast particles, which fly between cell walls overcoming the potential well (1). Therefore now we determine the maximum velocity $v_1^*(t_1)$ of particles departing from a cell wall in some moment $t_1$, when given particles still may be captured in the electromagnetic trap. Corresponding analytical calculations may be carry out for adiabatic increase of the potential well (1), when a change of the frequency $\omega(t)$ is negligible during the characteristic oscillation period $2\pi/\omega(t)$ of a trapped particle, that is

$$\frac{2\pi}{\omega(t)} \frac{d\omega(t)}{dt} \ll \omega(t). \qquad (4)$$

Then, for any trapped particle in our one-dimensional model, the following adiabatic invariant is constant [4]:

$$I(t) = (m/2\pi) \int dv\, dx, \qquad (5)$$

where the two-dimensional integral is taken over the coordinate $x$ and velocity $v$ of a moving particle. Phase trajectory equation of such a particle for the potential well (1) has the known form:

$$0.5 \, m \cdot v^2 + 0.5 \, m \cdot \omega(t)^2 \cdot x^2 = E(t, t_1), \qquad (6)$$

where $E(t, t_1)$ is the energy in the moment $t$ for a particle, which is captured by the trap after reflection from any cell wall (with the coordinate $x = l$ or $-l$) in the moment $t_1 < t$. Then we receive the following adiabatic invariant (5) for such particles:

$$I = E(t, t_1)/\omega(t). \qquad (7)$$

According to the invariant (7), the simple connection takes place between particle coordinate $x = l$ or $(-l)$ and its velocity $v_1$ in the moment $t_1$ of its reflection from a cell wall with corresponding values $x(t)$ and $v(t)$ in following moments $t > t_1$ of particle motion in the trap:

$$v(t)^2 + \omega(t)^2 \cdot x(t)^2 = [\omega(t)/\omega(t_1)] \cdot v_1(t_1)^2 + \omega(t_1) \cdot \omega(t) \cdot l^2. \qquad (8)$$

At the adiabatic change of the frequency $\omega(t)$ (4), the first turn of a trapped particle (for example in the point $z$ in Fig.1) after reflection from a wall in the moment $t_1$ occurs approximately in the moment $[t_1 + 0.5\pi/\omega(t_1)]$. The limitary initial velocity $v_1^*(t_1)$ for particles trapping is determined from the condition that such a particle touches the opposite wall of the cell (for example with the coordinate $x = l$ in Fig.1) in the indicated moment $[t_1 + 0.5\pi/\omega(t_1)]$. Thus, after substitution of values $v = 0$, $x^2 = l^2$ and $t = [t_1 + 0.5\pi/\omega(t_1)]$ in Eq.(8), we receive the following limitary velocity $v_1^*(t_1)$:

$$v_1^*(t_1) = l \cdot \sqrt{\omega(t_1) \{\omega[t_1 + 0.5\pi/\omega(t_1)] - \omega(t_1)\}} \approx l \cdot \sqrt{0.5\pi \cdot \left(\frac{d\omega(t_1)}{dt_1}\right)}. \qquad (9)$$

We note that, in case of the stationary potential well (1), it is impossible to capture particles in such a trap because they will fly between walls of the cell even when their initial departure velocity (from a wall) is close to zero. It is confirmed also by the formula (9) where the velocity $v_1^* = 0$ for the constant value $\omega$.

During sufficiently slow growth even a highly shallow potential well (1), it is possible to accumulate a large number of captured particles in this well if given particles are slowed up to velocities $v < v_1^*$ because of inelastic collisions with cell walls directly before their trapping. Indeed, let us analyze the case of the diffuse reflection of particles from a wall surface, after which the equilibrium (Maxwell) distribution $F(v)$ on particles velocities establishes [5]:

$$F(v) = \frac{1}{u\sqrt{\pi}} exp\left[-\left(\frac{v}{u}\right)^2\right], \qquad (10)$$

where $u = (2k_B T_w/m)^{0.5}$ is the most probable speed of free particles in the gas in the absence of the electromagnetic trap, $k_B$ is the Boltzmann constant and $T_w$ is the temperature of the cell walls. We will assume that the potential well (1) is so shallow that the relationship $\omega(t) \cdot l \ll u$ takes place and it is possible to neglect an electromagnetic trap influence on a distribution of comparatively fast particles which overcome the potential depth. Then the probability of capture of particles in the trap after their unitary collisions with cell walls in the moment $t$ is determined by the value:

$$p(t) = \int_0^{v_1^*(t)} F(v) dv \approx \frac{v_1^*(t)}{u\sqrt{\pi}}, \qquad (11)$$

where the limitary velocity $v_1^*(t)$ is determined by the formula (9) in case of the adiabatic growth of the potential well (1). A number of such trapped particles $N(t)$ in some cylindrical volume of our one-dimensional gas cell in the moment $t$ has the form:

$$N(t) = 2l \cdot S \cdot n \cdot \theta(t), \quad (12)$$

where $S$ is the area of the corresponding part of cell walls in the selected volume, $n$ is the equilibrium density of free particles in the absence of the electromagnetic trap, $\theta(t) \leq 1$ is the fraction of trapped particles in the cell and $\theta(t = 0) = 0$ in the initial moment $t=0$ of the trapping process. We receive the following balance equation on the basis of relationships (11) and (12):

$$\frac{dN(t)}{dt} = J \cdot p(t) \cdot [1 - \theta(t)] \cdot n \cdot S, \quad (13)$$

where $J = 2\int_0^\infty F(v) \cdot v \cdot dv = 0.564\, u$, and $J \cdot [1 - \theta(t)] \cdot n \cdot S$ is the flow of still nontrapped particles, which falls on the indicated part of cell walls with the area $S$ in the moment $t$. Then from Eq. (11)-(13) we receive the fraction $\theta(t)$ of particles in the cell, which are captured in the electromagnetic trap up to the moment $t$:

$$\theta(t) = 1 - exp\left[-\frac{0.282}{l \cdot \sqrt{\pi}} \int_0^t v_1^*(t_1) d\, t_1\right]. \quad (14)$$

In particular, for the linear time dependence (3) of the frequency $\omega(t) = (\omega_0 + \sigma \cdot t)$, the limitary velocity $v_1^* = l \cdot \sqrt{0.5\pi \cdot \sigma}$ (9) is constant and the function $\theta(t)$ (14) has the form:

$$\theta(t) = 1 - exp[-0.282 \cdot \sqrt{0.5\sigma} \cdot t]. \quad (15)$$

It is important also to analyze the following average energy $E_a(t)$ of particles captured in the electromagnetic trap up to the moment $t$:

$$E_a(t) = \frac{1}{\theta(t)} \int_0^t E(t, t_1) \frac{d\theta(t_1)}{dt_1} dt_1. \quad (16)$$

The energy $E(t, t_1)$ is determined by the relationship (6) and under our conditions may be presented in the following form on the basis of the adiabatic invariant (7):

$$E(t, t_1) = \frac{m \cdot \omega(t) \cdot v_1^2}{2\omega(t_1)} + \frac{m \cdot \omega(t_1) \cdot \omega(t) \cdot l^2}{2} \approx \frac{m \cdot \omega(t_1) \cdot \omega(t) \cdot l^2}{2}, \quad (17)$$

where we used the relationship $v_1^2 \ll \omega(t_1)^2 \cdot l^2$, which proceeds from the formula (9) at the adiabatic change of the frequency $\omega(t)$ (4). According to Eqs. (16) and (17), the initial average energy of trapped particles $E_a(t) \to 0.5m \cdot [\omega(t = 0) \cdot l]^2$ when $t \to 0$.

Fig.2 presents time dependences of the relative depth $[\omega(t)^2/\omega_0^2]$ of the potential well (1) and also the fraction $\theta(t)$ (15) and the average energy $E_a(t)$ (16) of particles captured in the trap at the linear growth of the frequency $\omega(t)$ (3). We can see that most of particles in the cell are trapped during the indicated period (Fig.2b), while the potential depth increases only on the factor 4 (Fig.2a). It is important, that in spite of growth approximately on the factor 3 during this time, the average energy $E_a(t)$ of given trapped particles still are much less that the mean kinetic energy $0.25\ m \cdot u^2$ of free particles without electromagnetic trap (Fig.2c). Such a "cooling" of trapped particles is caused by their inelastic collisions with cell walls, when their speeds may lower up to magnitudes available for following capture of these particles. The considered process may occur sufficiently quickly. For example, in Fig.2 the indicated time interval $10^4$ ($l/u$)~1s when the length of the one-dimensional cell $l$~1cm and the most probable speed of free particles $u$~$10^2$ m/s.

## 3. Discussion of results

Results of the previous section were obtained for the comparatively simple one-dimensional model of the gas cell and the definite potential well (1). Corresponding calculations are much more complicated for real two- and three-dimensional systems. However numerical calculations carried out by author for gas cells with cylindrical and spherical symmetry and various spatial dependences of increasing (in time) potential wells confirmed following qualitative results 1-4 obtained in the section 2.

*1). Even a highly shallow but increasing with time potential well will continuously capture sufficiently slow-speed particles of a strongly rarefied gas in a cell.*

*2). Such trapped particles will remain in the potential well and will not collide with cell walls even after going out of the corresponding nondecreasing electromagnetic field on a stationary value.*

*3). These electromagnetic traps are especially effective in cases of inelastic collisions of still nontrapped particles with walls of the gas cell. Then necessary slowdown of given particles is possible for their following capture in the potential well. The accumulation process of comparatively slow-speed particles in the trap takes place during intensification of the electromagnetic field because a probability of such slowdown collisions increases with time.*

*4). At definite conditions, an average speed of particles captured in considered electromagnetic traps will be much less than the most probable speed of such free particles in the gas cell without given traps.*

Of course, given results 1-4 are valid only in the absence of an interaction between particles in the gas cell. However such an interaction may be essential at a sufficiently high concentration of captured particles in a comparatively small volume of the trap.

We have considered new electromagnetic traps for structureless particles having an electric or magnetic moment. In practice such situations may take place, in particular, for a collection of micro- or nano-particles, which fly in a cell under conditions of the ultrahigh vacuum at action of the controllable nonhomogeneous electric (magnetic) field or laser radiation.

For analysis of possible capture of atoms and molecules in proposed traps, consideration of their quantum structure is necessary. Meanwhile, in definite cases, results obtained in this paper may be generalized also on such quantum objects. Thus, for example, it is possible creation of two- and three-dimensional traps for atoms and molecules by the nonhomogeneous laser radiation with frequencies essentially detuned from resonances with atomic (molecular) transitions [6-8]. Then the gradient force acts on atoms (molecules) in the direction to the point of minimum of the light induced potential well. However such a well is usually so shallow that a preliminary slowdown of particles up to very low speeds is necessary for their trapping. Unlike some atoms [1,6], such an effective "cooling" can not be realized for most molecules because of their complicated structure [7,8]. However it is possible to realize the necessary slowdown and following trapping of a large number of ground state molecules in a compact gas cell by the more universal method proposed in the present paper. Indeed, then an inelastic (in particular) diffuse reflection of nontrapped molecules from cell walls is necessary. A collection of captured molecules by the proposed method will have much less characteristic speed in comparison with free molecules in the absence of the electromagnetic trap. Thus, in particular, an essential decrease of Doppler widths of absorption (or fluorescence) spectral lines of trapped molecules in the gas cell may be recorded by the additional probe radiation.

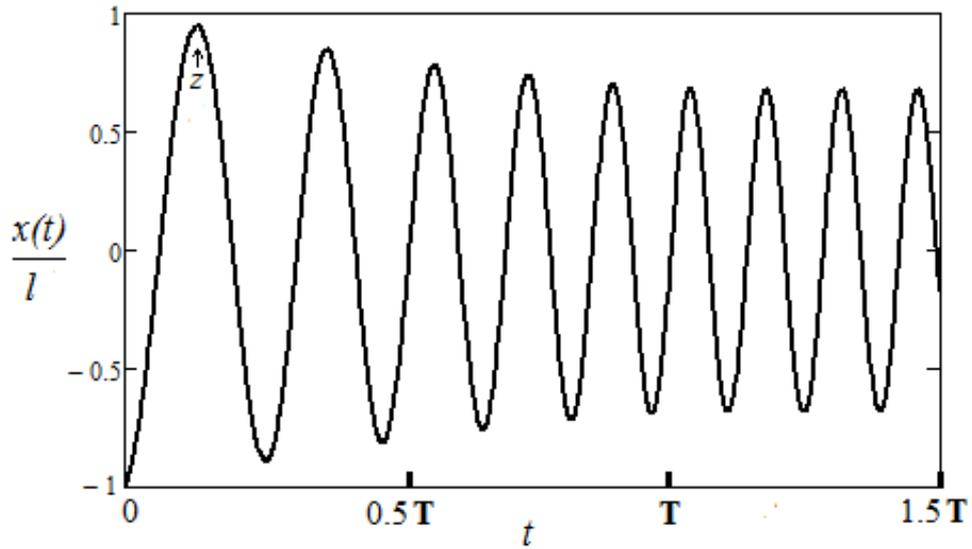

**Fig.1.** The one-dimensional motion of a particle, which is reflected with the velocity $v_1$ from the cell wall with the coordinate $x = -l$ in the moment $t=0$ in case of the time change of the frequency $\omega(t)$ according to the formula (3), when $v_1 = 5(l/T)$, $\omega_0 = 20/T$ and $\sigma = 25/T^2$.

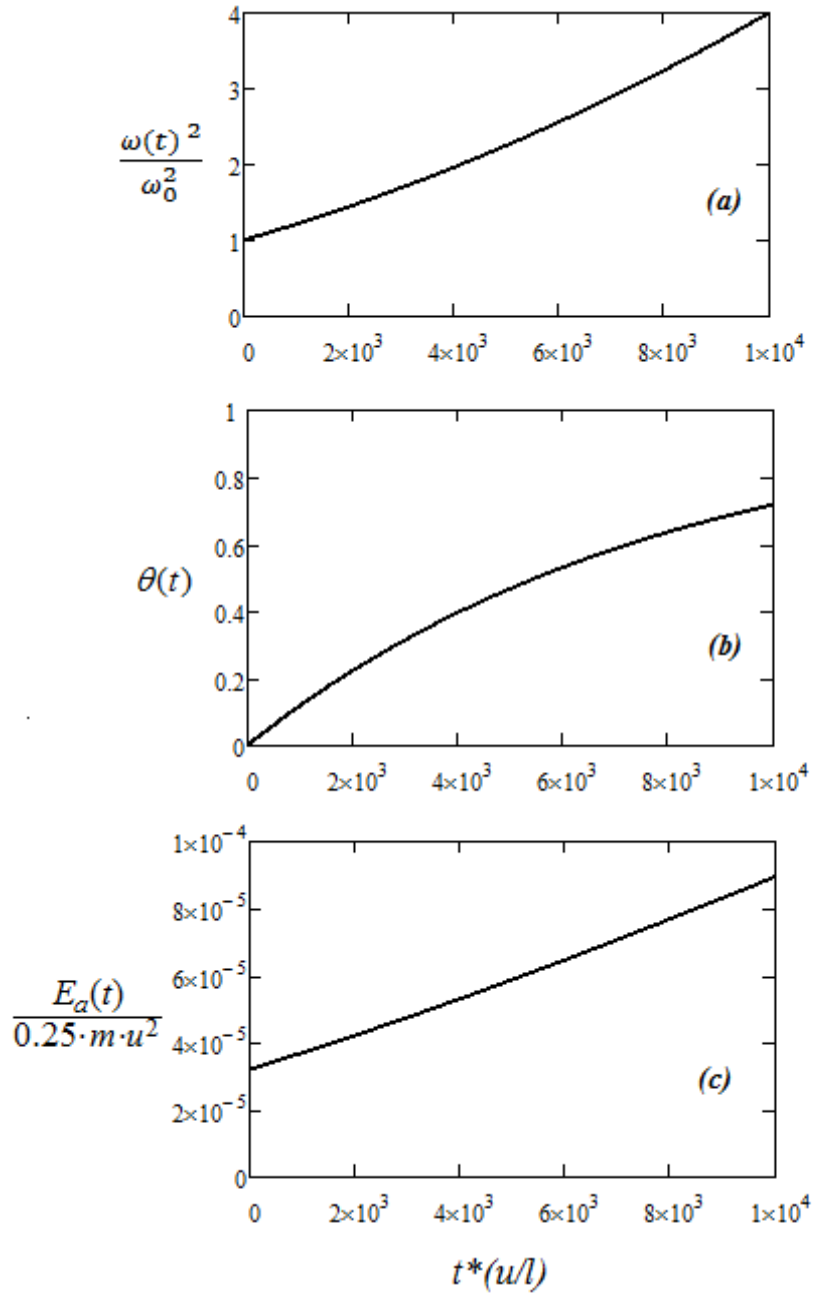

**Fig.2.** Time dependences of the relative potential depth~$[\omega(t)^2/\omega_0^2]$ (a) and also the fraction $\theta(t)$ (b) and the average energy $E_a(t)$ (c) of particles captured in the trap in case of the linear growth of the frequency $\omega(t) = \omega_0 + \sigma \cdot t$, when $\omega_0 = 4 \cdot 10^{-3}$ $(u/l)$ and $\sigma = 4 \cdot 10^{-7}$ $(u/l)^2$.